\documentstyle[12pt,dina4p]{article}
\begin{document}
\begin{titlepage}

\begin{flushright}
DESY 95--112\\
hep-th/9510011
\end{flushright}

\vspace{1cm}

\begin{center}

{\LARGE \sc Non--Commutative Geometry on \\
 [5mm]
 Quantum Phase--Space}

\vspace{1cm}

{\large M.\ Reuter \\
\vspace{1cm}
\noindent
Deutsches Elektronen--Synchrotron DESY \\
Notkestrasse 85 \\
D--22603 Hamburg \\
Germany
}
\end{center}


\vspace*{1cm}
 \begin{abstract}
A non--commutative analogue of the classical differential
forms is constructed on the phase--space of an arbitrary
quantum system. The non--commutative forms are universal
and are related to the quantum mechanical dynamics in the
same way as the classical forms are related to classical dynamics.
 They are constructed by applying the Weyl--Wigner symbol map
to the differential envelope of the linear operators on the
quantum mechanical Hilbert space. This leads to a representation
of the non--commutative forms considered by A.~Connes in terms
of multiscalar functions on the classical phase--space. In an
appropriate coincidence limit they define a quantum deformation of
the classical tensor fields and both commutative and
non--commutative forms can be studied in a unified framework.
We interprete the quantum differential forms in physical terms
and comment on possible applications.
 \end{abstract}
\end{titlepage}
%
%
\renewcommand{\theequation}{1.\arabic{equation}}
\setcounter{equation}{0}

 \section{Introduction}
 Non--commutative geometry is a fascinating new field
with a wide range of possible applications in physics and mathematics.
Among many other developments, both the
approach of Connes \cite{co}
          and of Dubois--Violette \cite{dbv}  were
used to construct particle physics models \cite{mco,mbdv}
        and  unified models including gravity
\cite{grav}. Also the scheme of
Coquereaux et al. \cite{coq} has been worked out in detail
\cite{mainz}.
Though one of the main inspirations of non--commutative geometry
are the operator algebras of quantum mechanics, whose
non--commutative nature is due to the non--commutativity of the
{\it phase--space} variables $p$ and $q$, it is space--time
which is turned into a non--commutative manifold in these papers.
A similar remark also applies to most of the work done on
non--commutative geometries based on quantum groups
\cite{wz, qpl}.

In the present paper we are instead interested in the non--commutative
geometry on the phase--space of an arbitrary quantum mechanical
system with $N$ degrees of freedom. As we shall see, there is
a very natural way of introducing a non--commutative version of
differential forms on phase--space. The resulting generalized
$p$--forms have very interesting properties both from a
geometrical and a dynamical point of view. Our approach has
two basic ingredients:

(I) Given an arbitrary algebra $A$, non--commutative geometry
\cite{co, review}
             tells us how to associate the algebra $\Omega A$
of universal differential forms to it. One of the possible
constructions of $\Omega A$ proceeds by identifying the
universal $p$--forms with special elements of the
$(p + 1)$--fold tensor product $A \otimes A \otimes \cdots
                                       \otimes A$.
For $A$ we shall take the algebra of linear operators
(observables, unitary transformations, etc.) acting on the
quantum mechanical Hilbert space ${\cal H}$ of the system under
consideration. The universal $p$--forms  are operators  on
$ {\cal H} \otimes {\cal H} \otimes \cdots \otimes {\cal H}$ then.

(II) Following the ideas of what is known as deformation
or star--product quantization \cite{lich} we describe the
quantum system at hand not by operators and Hilbert space
vectors but rather by functions over the classical phase
space, henceforth denoted ${\cal M}_{2N}$. In this manner the
vector $\psi \in {\cal H}$ and the operator $\hat{a} \in A$
become replaced
by the Wigner function $W_\psi$ and the operator symbol
$a \equiv {\rm symb}~(\hat{a})$, respectively. Both
$W_\psi \equiv W_\psi (\phi^a)$ and $a \equiv a (\phi^a)$
are scalar c--number functions over ${\cal M}_{2N}$ whose
local coordinates are denoted as $\phi^a$.
Similarly the non--commutative $p$--forms in
$A \otimes \cdots \otimes A$ are turned into functions
$F_p (\phi_0, \phi_1, \cdots , \phi_p)$ which depend on
$p + 1$ phase--space points. We shall refer to them
as multiscalars.

The multiscalars provide a link between
the classical and the non--\linebreak commutative tensor calculus.
At the classical level $(\hbar = 0)$ the functions
$F_p (\phi_0, \phi_1,  \linebreak     \cdots, \phi_p)$ define tensor
fields (and in particular differential forms) if
we perform the coincidence limit where all
points $\phi_i$ are very ``close'' to each other.
Setting $\phi^a_i = \phi^a_0 + \eta^a_i, i = 1, \cdots, p$,
and expanding to first order in $\eta_i$ yields
terms of the type $\partial\,^{(1)}_{a_1} \cdots \partial\,
                                         ^{(p)}_{a_p}
F_p \, (\phi_0, \phi_0, \cdots , \phi_0) \cdot \eta^{a_1}_1
\eta^{a_2}_2 \cdots \eta^{a_p}_p$, where $ \partial^{(i)}_a$ is the
derivative with respect to $\phi^a_i$. All derivatives act
on different arguments of $F_p$. Therefore the coefficients
of $\eta^{a_1}_1 \cdots \eta^{a_p}_p$ define a classical $p$--form
upon antisymmetrization. In the non--commutative case
$(\hbar > 0)$ it is well known \cite{review} that the abstract
construction of $\Omega^p A$ has a concrete realization in
terms of functions of $p + 1$ arguments which vanish if two
neighboring arguments are equal. The pseudoscalars $F_p$
resulting from the construction (II) are very similar
to these functions. For $\hbar = 0$ they, too, vanish if two
adjacent arguments are equal, and for $\hbar > 0$
they satisfy a deformed version of this condition.

The main virtue of our deformation theory approach is that
it represents the non--commutative $p$--forms in the
same setting as classical $p$--forms, namely as c--number
functions on the classical phase--space. The deformation
parameter $\hbar$ allows for a smooth interpolation between
the classical and the non--commutative case.

In order to fix our notation and to collect some formulas
which we shall need later on let us recall some elements
of the phase--space formulation of quantum mechanics
\cite{lich, mo, ber, lj}. We consider a set of operators
$\hat{f}, \hat{g}, \cdots$ on some Hilbert space ${\cal H}$, and
we set up a one--to--one correspondence between these operators
and the complex--valued functions $f, g, \cdots \in
{\rm Fun}({\cal M})$ defined over a suitable manifold ${\cal M}$.
                                                           We write
$ f = {\rm symb}~(\hat{f})$,
                       and we refer to the function $f$ as the
symbol of the operator $\hat{f}$. The symbol map
``symb'' is linear and has a well--defined inverse. An
important notion is the ``star product'' which implements the
operator multiplication at the level of symbols:

\begin{equation}
{\rm symb}~(\hat{f} \,\, \hat{g}) = {\rm symb}~(\hat{f})~\ast~
                {\rm symb}~(\hat{g})
\end{equation}
The star product is non--commutative, but associative, because
``{\rm symb}''
    provides an algebra homomorphism  between the operator
algebra and the symbols. In the physical applications we have
in mind, the Hilbert space ${\cal H}$ is the state space of a
quantum mechanical system, and the manifold
${\cal M} \equiv {\cal M}_{2N}$
                                                   is the
 $2N$--dimensional classical phase--space of this system.
 Quantum mechanical operators $\hat{f}$ are then represented
 by functions $f = f (\phi)$, where $ \phi^a =
 (p^1, \cdots, p^N, q^1, \cdots, q^N), a = 1, \cdots,
 2 N$ are canonical coordinates on ${\cal M}_{2N}$.
 Here and in the following we assume that the phase--space
 has the topology of ${\bf R}^{2N}$. Hence, by Darboux's theorem,
 we may assume that the symplectic 2--form of ${\cal M}_{2N}$,
 $\omega = \frac{1}{2} \omega_{ab} d \phi^a \wedge d \phi^b$, has
constant components:
\begin{equation}
\omega_{ab}   =  \left ( \begin{array}{cc}
                                        0  & I_N \\
-I_N & 0 \end{array} \right )
\end{equation}
The inverse of this matrix,
\begin{equation}
\omega^{ab} = \left ( \begin{array}{cc}
0 & -I_N \\
I_N & 0 \end{array} \right )
\end{equation}
defines the Poisson bracket
\begin{equation}
\{ f, g \}_{\rm pb} (\phi) =
\partial_a f (\phi) \omega^{ab} \partial_b g (\phi)
\end{equation}
where $\partial_a \equiv \frac{\partial}{\partial \phi^a}$.
In the language of quantum mechanics, specifying the symbol
map means fixing an ordering prescription, because it
associates a unique operator $\hat{f} (\hat{p}, \hat{q}) =
{\rm symb}^{-1} (f (p, q))$ to any classical phase function
$f (p,q)$. A typical example is the Weyl symbol. It
associates the symmetrically, or Weyl--ordered operator
$\hat{f} (\hat{p}, \hat{q})$ to the polynomial $f (p, q)$.
For instance, ${\rm symb}^{-1} (p q) = \frac{1}{2} (\hat{p}
\hat{q} + \hat{q} \hat{p})$. Formally the Weyl symbol $f$
of the operator $\hat{f}$ is given by
\begin{equation}
f (\phi^a) = \int \frac{d^{2N} \phi_0}{(2 \pi \hbar)^N}
\, \exp \, \left [ \frac{i}{\hbar} \phi^a_0 \omega_{ab} \phi^b \right ]
{\rm Tr} \left [ \widehat{T} (\phi_0) \hat{f} \right ]
\end{equation}
with the operators
\begin{equation}
\widehat{T} (\phi_0) = \, \exp \, \left [ \frac{i}{\hbar} \phi^a_0
\omega_{ab} \hat{\phi}^b \right ] = \exp \left [ \frac{i}{\hbar}
(p_0 \hat{q} - q_0 \hat{p})\right ]
\end{equation}
which generate translations on phase--space.\\

For any pair of Weyl symbols, $f, g \in {\rm Fun} ({\cal M}_{2N})$,
the star product reads
\begin{eqnarray}
(f \ast g) (\phi) = f (\phi) \exp \left [ i \frac{\hbar}{2}
\stackrel{\leftarrow}{\partial_a} {\omega}^{ab} \stackrel{\rightarrow}
{\partial_b} \right ] g (\phi) \nonumber\\
\equiv \exp \left [ i \frac{\hbar}{2} \omega^{ab} \partial_a^{(1)}
\partial_b^{(2)} \right ] f (\phi_1) g (\phi_2)
\Bigl|_{\phi_1  = \phi_2=\phi} \Bigr.
\end{eqnarray}
where $\partial_a^{(1,2)} \equiv \partial / \partial \phi^a_{1,2}$.
In the classical limit $\hbar \rightarrow 0$ one has $(f \ast
g) (\phi) = f (\phi) g (\phi) + 0 (\hbar)$, i.e. the star
product is a ``deformation'' of the pointwise product of two
functions. The Moyal bracket of two symbols is defined by
\begin{equation}
\{ f, g \}_{\rm mb} = \frac{1}{i \hbar}
(f \ast g - g \ast f) = {\rm symb}~\left(\frac{1}{i \hbar}
 [\hat{f}, \hat{g}]\right)
 \end{equation}

\noindent
In the classical limit it reduces to the Poisson bracket:
$\{ f, g \}_{\rm mb} = \{ f, g \}_{\rm pb}
+ O (\hbar^2)$. The most remarkable property of the
star product is its associativity. As a consequence
of it, the Moyal bracket obeys the Jacobi identity.

Any density matrix $\hat{\rho} (t)$ time--evolves
according to von Neumann's equation
\begin{equation}
\label{1.9}
i \hbar \,\, \partial_t \hat{\rho} = \left [ \hat{H}, \hat{\rho}
\right ]
\end{equation}
Applying the symbol map to both sides of this equation
we obtain Moyal's equation for the ``pseudodensity''
$\rho = {\rm symb} (\hat{\rho})$:
\begin{equation}
\label{1.10}
\partial_t \rho (\phi) = \{ H, \rho \}_{\rm mb}
 \equiv V_H (\phi) \rho (\phi)
 \end{equation}
 In the classical limit, (\ref{1.10}) reduces to Liouville's
 equation $\partial_t \rho = \{ H, \rho \}_{\rm pb}$.
 In eq. (\ref{1.10}) we have introduced the pseudo--differential
 operator \cite{fz, winf}
 \begin{equation}
\label{1.11}
 V_H (\phi) = \frac{2}{\hbar} H (\phi) \sin
 \left [ \frac{\hbar}{2} \stackrel{\leftarrow}{\partial_a}
 \omega^{ab} \stackrel{\rightarrow}{\partial_b}\right ]
 \end{equation}
 which reduces in the classical limit to minus the hamiltonian
 vector field: $V_H = - \omega^{ba} \partial_a H \partial_b + O
 (\hbar^2)$. The operators (\ref{1.11}) form a closed algebra:
 \begin{equation}
 \label{1.12}
 \left [ V_{H_1}, V_{H_2} \right ] = V_{\{ H_1, H_2 \}_{\rm mb}}
 \end{equation}
 This relation is most easily established by noting that, at the
 operatorial level, $V_H$ corresponds to
 $(L_H - R_H) / i \hbar$ where $L_H (R_H)$ denotes the
 left(right) multiplication by $\hat{H}$. (See \cite{winf} for
 details).

 The remaining sections of this paper are organized as follows.
 In section~2 we introduce the multiscalar functions $F_p$, define
 an exterior and a Lie derivative for them, and establish
 their relation to classical differential forms. In section 3
 we provide some details about the non--commutative universal
 differential forms which will be needed in order to relate the
 abstract algebraic construction to the multiscalars. In
 section 4 the symbol map is applied to the operatorial
 construction of $\Omega A$, and the quantum mechanical
 multiscalar formalism is obtained. In section 5 the coincidence
 limit of the quantum mechanical multiscalars is investigated,
 and the physical meaning of the quantum deformation from
 classical to non--commutative forms is discussed.

\renewcommand{\theequation}{2.\arabic{equation}}
\setcounter{equation}{0}

\section{Classical Differential Forms from
                               \protect \newline Multiscalars}

In this section we construct the exterior algebra of
antisymmetric tensor fields by starting from
multiscalar functions defined on the manifold under
consideration. This section serves as a preparation
for an analogous treatment of the quantum deformed case.

Let $\bigwedge^p_{\rm MS}$ denote the set of ``multiscalar"
functions $F_p (\phi_0, \phi_1, \cdots, \phi_p)$ which
depend on $p + 1$ arguments $\phi_i, i = 0, 1, \cdots,
p$ and which vanish whenever two neighboring arguments are
equal \cite{review}:
\begin{equation}
\label{2.1}
F_p \left (\phi_0, \phi_1, \cdots, \phi_{i-1}, \phi_i, \phi_i,
\phi_{i+2}, \cdots, \phi_p \right ) = 0
\end{equation}

\noindent
Now we define a map
\begin{displaymath}
\delta : \, \, \bigwedge \nolimits^p_{\rm MS} \longrightarrow
\bigwedge \nolimits^{p+1}_{\rm MS}
\end{displaymath}

\noindent
by
\begin{equation}
\label{2.2}
\left ( \delta F_p \right ) \left ( \phi_0, \cdots, \phi_{p+1} \right )
= \sum_{i = 0}^{p+1} (-1)^i \, F_p \left ( \phi_0, \cdots,
\phi_{i-1}, \widehat{\phi}_i, \phi_{i+1}, \cdots, \phi_{p+1} \right)
\end{equation}

\noindent
where the caret over $\phi_i$ means that this argument
is omitted. The $\delta$--operation maps a function of
$p+1$ arguments onto a function of $p+2$ arguments.
It is easy to verify that the image $\delta F_p$ is in
$\bigwedge^{p+1}_{\rm MS}$, i.e.  that it vanishes if
two adjacent arguments are equal. The first few
examples are
\begin{eqnarray}
\left( \delta F_0 \right) \left(\phi_0, \phi_1 \right)
& = & F_0 \left( \phi_1 \right)  - F_0 \left(\phi_0\right)
\label{2.3} \\
\left( \delta F_1 \right) \left(\phi_0, \phi_1, \phi_2 \right)
& = & F_1  \left( \phi_1, \phi_2 \right)
 - F_1 \left(\phi_0, \phi_2 \right) + F_1 \left(\phi_0, \phi_1
 \right) \label{2.4} \\
\left( \delta F_2 \right) \left(\phi_0, \phi_1, \phi_2, \phi_3\right)
& = & F_2  \left( \phi_1, \phi_2, \phi_3 \right)
 - F_2 \left(\phi_0, \phi_2, \phi_3 \right)  \nonumber \\
 \hspace*{2cm}
 & &      + F_2 \left(\phi_0, \phi_1, \phi_3 \right) - F_2 \left(\phi_0,
 \phi_1, \phi_2 \right)   \label{2.5}
\end{eqnarray}

\noindent
Remarkably enough, $\delta$ turns out to be nilpotent:
\begin{equation}
\label{2.6}
\delta^2 = 0
\end{equation}

\noindent
This can be checked explicitly, but it is most easily seen by
noting that if we interpret $F_p \left( \phi_0, \cdots, \phi_p \right)$
as a $p$--simplex with vertices $\phi_0, \cdots, \phi_p$ then
$\delta$ acts exactly like the nilpotent boundary operator of
conventional simplicial homology \cite{naka}.

Henceforth we shall refer to a function $F_p \in \bigwedge^p_{\rm MS} $
as a ``$p$--form''. On the direct sum
\[
\bigwedge \nolimits^\ast_{\rm MS} \, = \, \bigoplus^\infty_{p = 0} \,
\bigwedge \nolimits^p_{\rm MS}
\]

\noindent
there exists a natural product of a $p$--form $F_p$ with
a $q$--form $G_q$ yielding a $(p + q)$--form $F_p \bullet G_q$:
\begin{eqnarray}
\label{2.7}
\left( F_p \bullet G_q \right) \left(\phi_0, \phi_1,
\cdots, \phi_p, \phi_{p+1}, \cdots, \phi_{p+q} \right) \nonumber\\
= F_p \left(\phi_0, \phi_1, \cdots, \phi_p \right) G_q \left(\phi_p,
\phi_{p+1}, \cdots, \phi_{p+q} \right)
\end{eqnarray}

For later use we note that $\delta$  obeys a kind of Leibniz rule
with respect to this product:
\begin{equation}
\label{2.8}
\delta \left(F_0 \bullet G_0 \right) = \left(\delta F_0 \right)
\bullet G_0 + F_0 \bullet \left(\delta G_0 \right)
\end{equation}

\noindent
This relation is easily proven by using (\ref{2.3}). It has no
analogue for higher $p$--forms.     \\

\vspace*{5mm}

\par \noindent
{\large Multiscalars}  \\

Up to now we have not yet specified the precise nature of the
functions $F_p$ and their arguments $\phi_i$. From now
on we assume that the $\phi_i \equiv \left(\phi_i^a \right)$
are local coordinates on some manifold ${\cal M}$, and that the
$F_p$'s are smooth functions which transform as multiscalars
under general coordinate transformations (diffeomorphisms)
on ${\cal M}$. This means in particular that $F_p$ evolves
under the flow generated by some vector field
$h = h^a (\phi) \partial_a, \partial_a \equiv \frac
{\partial}{\partial \phi^a}$,
according to

\begin{equation}
\label{2.9}
\partial_t  F_p  \left( \phi_0, \cdots, \phi_p; t \right)
= \sum_{i = 0}^{p} {\rm V}  \left( \phi_i \right)
F_p \left(\phi_0, \cdots, \phi_p; t \right)
\end{equation}
\noindent
where
\begin{equation}
\label{2.10}
{\rm V}  \left(\phi_i \right) = - h^a \left(\phi_i \right)
\partial_a^{(i)} \equiv - h^a \left(\phi_i \right)
\frac{\partial}{\partial \phi_i^a}
\end{equation}

\noindent
acts only on the $i-{\rm th}$ argument of $F_p$. Here the
``time'' $t$ parametrizes points along the flow lines
of the vector field $h$.

In this paper we restrict our attention to symplectic manifolds
${\cal M} \equiv {\cal M}_{2N} = {\bf R}^{2N}$ and to
vector fields which are hamiltonian, i.e. we assume that
(locally)
\begin{equation}
\label{2.11}
h^a (\phi) = \omega^{ab} \partial_b H (\phi)
\end{equation}

\noindent
for some generating function (``Hamiltonian'') $H$.
Let us introduce the operators
\begin{equation}
\label{2.12}
{\cal L}_p [H] = - \sum_{i = 0}^{p} {\rm V}_H (\phi_i) ,
\end{equation}

\begin{equation}
\label{2.13}
{\rm V}_H (\phi_i) \equiv \omega^{ab}
\partial_a H (\phi_i) \frac{\partial}{\partial \phi_i^b}
\end{equation}

\noindent
and let us determine their commutation relations for
different generating functions. Using
\begin{equation}
\label{2.14}
\left[ {\rm V}_{H_1} (\phi_i), {\rm V}_{H_2} (\phi_j) \right]
= \delta_{ij} {\rm V}_{{\{H_1, H_2\}}_{\rm pb}} (\phi_i)
\end{equation}

\noindent
we see that the ${\cal L}_p$'s form a closed algebra:
\begin{equation}
\label{2.15}
\left[ {\cal L}_p \left[H_1 \right],
         {\cal L}_p \left[ H_2 \right] \right]
= - {\cal L}_p \left[ \{H_1, H_2 \}_{\rm pb} \right]
\end{equation}

\noindent
This is the Lie algebra of infinitesimal symplectic
diffeomorphisms (canonical transformations) on
${\cal M}_{2N}$. It is therefore natural to look at
${\cal L}_p$ as the Lie derivative appropriate for the
generalized $p$--forms $F_p$. It is easy
to see that ${\cal L}_p$ commutes with the ``exterior
derivative'' on $\bigwedge \nolimits ^p_{\rm MS},
 \delta \equiv \delta_p$:
\begin{equation}
\label{2.16}
\delta_p {\cal L}_p = {\cal L}_{p+1} \, \delta_p
 \end{equation}

Later on we shall replace the generator ${\rm V}_H (\phi_i)$,
eq. (\ref{2.13}), by its quantum deformed (Moyal)
analogue (\ref{1.11}). Even then the relations
(\ref{2.15}) and (\ref{2.16}) remain valid, provided one replaces the
Poisson bracket in (\ref{2.15}) by the corresponding Moyal bracket.\\

\vspace*{5mm}

\par \noindent
{\it \large The coincidence limit}  \\

We are now going to show how in the limit when the
arguments of the multiscalar $F_p (\phi_0, \phi_1, \cdots,
\phi_p)$ are very ``close'' to each other, the generalized
$p$--form $F_p \in \bigwedge^p_{\rm MS}$ gives rise to
a conventional $p$--form. We set
\begin{eqnarray}
\label{2.17}
\phi^a_0 & = & \phi^a  \nonumber \\
\phi^a_i & = & \phi^a  + \eta^a_i \, , \, i = 1, \cdots, p
\end{eqnarray}

\noindent
and expand $ \left( \partial\,^{(i)}_a \equiv \partial / \partial
\phi^a_i \right)$
\begin{eqnarray}
\label{2.18}
F_p \left( \phi, \phi + \eta_1, \cdots, \phi + \eta_p \right)
= \exp \left[  \sum_{i = 1}^{p} \eta^a_i \partial_a^{(i)}
\right] F_p (\phi, \phi, \cdots, \phi)
\end{eqnarray}

\noindent
to lowest order in $\eta_1^a, \cdots, \eta^a_p$.
We keep only terms in which all $\eta_i$'s are different,
e.g. $\eta^a_1 \eta^b_2$, but not $\eta^a_1 \eta^b_1$, say.
In this manner we obtain a sum of terms of the type
\begin{equation}
\label{2.19}
\eta^{a_1}_{i_1} \eta^{a_2}_{i_2} \cdots \eta^{a_l}_{i_l}
 \, \, \partial^{(i_1)}_{a_1} \cdots \partial^{(i_l)}_{a_l}
 F_p (\phi, \phi, \cdots, \phi)
 \end{equation}

\noindent
for $0 \le l \le p$. An important, though trivial, observation
is that, after ``stripping off'' the $\eta$'s, the quantities
$\partial^{(i_1)}_{a_1} \cdots \partial^{(i_l)}_{a_l}
F_p (\phi, \phi, \cdots, \phi)$, for $i_1, \cdots, i_l$ fixed,
transform as the components of a covariant tensor field of rank
$l$, because on each $\phi$--argument there acts at most one
derivative. Upon explicit antisymmetrization in
the indices $a_1, \cdots, a_l$ we obtain the components of
an $l$--form:
\[
\partial^{(i_1)}_{\left[ a_1 \right.} \cdots \partial^{(i_l)}
_{\left. a_l \right]} F_p \, (\phi, \phi, \cdots, \phi)
\]

These remarks also apply to any generic multiscalar $F_p$.
What is special about the generalized forms $F_p \in
\wedge^p_{\rm MS}$ is that for them the
expansion of (\ref{2.18}) does not contain any term with
$ l < p$, but only the one with the maximal rank $l = p$:
\begin{eqnarray}
\label{2.20}
F_p  \left( \phi, \phi + \eta_1, \cdots, \phi + \eta_p \right)
  \hspace*{45mm}\nonumber \\
 = \eta^{a_1}_1 \eta^{a_2}_2 \cdots  \eta^{a_p}_p
\, \, \partial^{(1)}_{a_1} \cdots \partial^{(p)}_{a_p}
F_p (\phi, \phi, \cdots, \phi) + O \left(\eta^2_i \right)
\end{eqnarray}

\noindent
In appendix A we show that eq. (\ref{2.20}) follows from the fact
that $F_p$ vanishes if two neighboring arguments coincide.

It is convenient to look at the ordinary differential
form induced by a certain multiscalar
$F_p \in \bigwedge^p_{\rm MS} \equiv \bigwedge^p_{\rm MS}
\left( {\cal M}_{2N} \right)$ as the image of the so-called
``classical map'' \cite{review}
\[
{\rm Cl}: \bigwedge \nolimits ^p_{\rm MS} \left({\cal M}_{2N}\right)
\longrightarrow \bigwedge \nolimits ^p \left({\cal M}_{2N} \right)
\]

\noindent
which is defined by

\begin{equation}
\label{2.21}
\left[ {\rm Cl} \left( F_p \right) \right]
(\phi) = \partial^{(1)}_{a_1} \cdots
 \partial^{(p)}_{a_p} F (\phi, \cdots, \phi)
\, d \phi^{a_1} \wedge \cdots \wedge d \phi^{a_p}
\end{equation}

\noindent
Under the Cl--map the $\bullet$--product (\ref{2.7}) becomes the
standard wedge product of the exterior algebra
$\bigwedge^\ast \left({\cal M}_{2N} \right)
= \bigoplus\limits^{2N}_{p = 0} \bigwedge^p \left({\cal M}_{2N}
\right)$:
\begin{equation}
\label{2.22}
{\rm Cl} \left( F_p \bullet G_q \right) = {\rm Cl} (F_p)
\wedge {\rm Cl} (G_q)
\end{equation}

\noindent
Similarly, the operator $\delta$ goes over into the exterior
derivative,
\begin{equation}
\label{2.23}
{\rm Cl} \left( \delta F_p \right) = d \, {\rm Cl} (F_p) \, ,
\end{equation}

\noindent
and ${\cal L}_p [H]$ of (\ref{2.12}) with (\ref{2.13}) becomes the Lie
derivative along $h$,
\begin{equation}
\label{2.24}
{\rm Cl} \left ( {\cal L}_p [H] F_p \right) = l_h
{\rm Cl} (F_p)
\end{equation}

\noindent
It acts on the components of $\alpha \in \bigwedge^p
\left( {\cal M}_{2N} \right)$ in the usual way:
\begin{eqnarray}
\label{2.25}
\begin{array}{ll}
l_{\displaystyle h} \alpha_{\displaystyle a_1 \cdots a_p}
   & = h^{\displaystyle b} \partial_{\displaystyle b}
\alpha_{\displaystyle a_1 \cdots a_p}
 + \partial_{\displaystyle a_1} h^{\displaystyle b}
\alpha_{\displaystyle b a_2 \cdots a_p}  \\
&  +  \partial_{\displaystyle a_2}
h^{\displaystyle b} \alpha_{\displaystyle a_1 b a_3 \cdots a_p}  +
\cdots
\end{array}
\end{eqnarray}

The proofs of eqs. (\ref{2.23}) and (\ref{2.24})
                                      can be found in appendix B.
Here we only illustrate (\ref{2.24}) for $p = 1$. In this case the
``Lie transport'' of the biscalar $F_1 \left( \phi_0, \phi_1 \right)$
 is described by the equation
\begin{eqnarray}
\label{2.26}
- \partial_t F_1 \left( \phi_0, \phi_1; t \right) & = & {\cal L}_1
F_1 \left( \phi_0, \phi_1; t \right) \\
& = & \left[ h^a \left(\phi_0 \right) \frac{\partial}{\partial
\phi^a_0} + h^a \left( \phi_1 \right)
\frac{\partial}{\partial \phi^a_1} \right] F_1
\left( \phi_0, \phi_1; t \right) \nonumber
\end{eqnarray}

\noindent
Now we set $\phi_0 = \phi, \phi_1 = \phi + \eta$, and compare
the coefficients of $(\eta)^0$ and $(\eta)^1$.
 By using $\frac{\partial}{\partial \phi^a}
F_1 (\phi, \phi; t) = 0$, with the derivative acting on both
arguments, one finds that the coefficient of
$(\eta)^0$ vanishes on both sides of the equation. The
result at order $(\eta)^1$ is
\begin{equation}
\label{2.27}
- \partial_t \alpha_a (\phi) = h^b \partial_b \alpha_a (\phi) +
\partial_a h^b \alpha_b (\phi) \equiv l_h \alpha_a
(\phi)
\end{equation}

\noindent
where
\begin{equation}
\label{2.28}
\alpha_a (\phi) = \frac{\partial}{\partial \phi_1^a}
F_1 \left( \phi_0, \phi_1 \right) \Bigl|_{ \phi_0 = \phi_1 = \phi}
= \left[ {\rm Cl} \left( F_1 \right) \right] (\phi) \Bigr.
\end{equation}

\newpage
\par \noindent
{\large A special class of multiscalars}\\

Let us assume we are given a set of scalar functions
on ${\cal M}_{2N}, f_i (\phi), i = 0, 1, \cdots, p$. Then
we can construct the following $F_p \in \bigwedge^p_{\rm MS}$
out of them:
\begin{eqnarray}
\label{2.29}
F_p \left( \phi_0, \phi_1, \cdots, \phi_p \right) =
f_0 \left(\phi_0 \right) \left[ f_1 \left( \phi_1 \right)
- f_1 \left( \phi_0 \right) \right] \nonumber \\
\left[ f_2 \left(\phi_2 \right) - f_2 \left( \phi_1 \right)
\right] \cdots \left[ f_p \left( \phi_p \right) -
f_p \left( \phi_{p-1} \right) \right]
\end{eqnarray}

\noindent
Generalized forms of this type will play an important role
later on. Inserting (\ref{2.17}) and expanding in $\eta$,
one finds that they have a particularly simple classical
limit:
\begin{eqnarray}
\label{2.30}
\left[ {\rm Cl} (F_p) \right] (\phi) = f_0 (\phi) d f_1
(\phi) \wedge d f_2 (\phi) \wedge \cdots \wedge
d f_p (\phi)
\end{eqnarray}

\noindent
By virtue of (\ref{2.23}) we know that
\begin{equation}
\label{2.31}
d \, {\rm Cl} (F_p) = d f_0 \wedge d f_1 \wedge \cdots \wedge
d f_p = {\rm Cl} (\delta F_p)
\end{equation}

\noindent
It is instructive to verify the second equality directly.
For $p = 1$, say, we can apply eq. (\ref{2.4}) to
$F_1 \left( \phi_0, \phi_1 \right) = f_0 \left( \phi_0
\right) \left[ f_1 \left(\phi_1 \right) - f_1 \left( \phi_0
\right) \right]$ and find
\begin{eqnarray}
\label{2.32}
\left( \delta F_1 \right) \left( \phi_0, \phi_1, \phi_2 \right) = \, \,
f_0 \left( \phi_1 \right) \left[ f_1 \left( \phi_2 \right) \right.
& - & \left. f_1 \left( \phi_1 \right) \right]  \nonumber \\
- f_0 \left( \phi_0 \right) \left[ f_1 \left( \phi_2 \right) \right.
& - & \left. f_1 \left( \phi_0 \right) \right] \nonumber \\
+ f_0 \left( \phi_0 \right) \left[ f_1 \left( \phi_1 \right) \right.
& - &  \left.  f_1 \left( \phi_0 \right)  \right] \nonumber\\
=  \left[ f_0 \left( \phi_1 \right) - f_0 \left( \phi_0 \right) \right]
\, \, \left[  \right. f_1 \left( \phi_2 \right) & - &  f_1 \left( \phi_1
 \right) \left. \right] ,
\end{eqnarray}

\noindent
as it should be. In the next section we shall introduce a set
of algebraic tools which render manipulations
of this type much more transparent.

\renewcommand{\theequation}{3.\arabic{equation}}
\setcounter{equation}{0}

\section{Non--Commutative Universal Differential \newline Forms}

Now we turn to a different subject whose relation to the
multiscalars of the previous section will become clear
later. In this section we briefly review some properties of the
universal differential forms in A.~Connes' non--commutative
geometry \cite{co}. We partly follow the presentation of ref.
\cite{review}.

To any algebra $A$ we can associate its universal differential
envelope $\Omega A$, the algebra of ``universal differential
forms''. Later on we shall identify $A$ with the linear operators
acting on a quantum mechanical Hilbert space, but for the time
being we make no assumptions about $A$. The construction of
$\Omega A$ proceeds as follows. To each element $a \in A$
we associate a new object $\delta a$. As a vector space,
$\Omega A$ is defined to be the linear space of words
built from the symbols $a_i \in A$ and $\delta a_i$, e.g.,
$a_1 \delta a_2 a_3 a_4 \delta a_3$. The multiplication
in $\Omega A$ is defined to be associative and distributive
over the addition $+$. The product of two elementary words
is obtained by simply concatenating the two factors. For
instance,
\[
\left( a_1 \delta a_2 \right) \cdot
\left( \delta a_3 a_4 \delta a_1 \right) = a_1 \delta a_2
\delta a_3 a_4 \delta a_1
\]
One imposes the following relation (a kind of Leibniz rule)
between the elements $a_1, a_2, \cdots \in A$ and
$\delta a_1, \delta a_2, \cdots \,\,$:
\begin{equation}
\label{3.1}
\delta \left( a_1 a_2 \right) = \left( \delta a_1 \right)
a_2 + a_1 \delta a_2
\end{equation}
By virtue of this relation, any element of $\Omega A$ can be
rewritten as a sum of monomials of the form
\begin{eqnarray}
a_0 \delta a_1 \delta a_2 \cdots \delta a_p \label{3.2}\\
\mbox{or} \hspace*{48mm} \nonumber \\
    \delta a_1 \delta a_2 \cdots \delta a_p\label{3.3}
\end{eqnarray}

\noindent
This form can be achieved by repeatedly applying the trick
\begin{equation}
\left( \delta a_1 \right) a_2 = \delta \left( a_1 a_2 \right)
- a_1 \delta a_2
\end{equation}

In order to put the two types of monomials, (\ref{3.2}) and
(\ref{3.3}), on an equal footing it is convenient to add a
new unit ``1'' to $A$, which is different from the unit
$A$ might have had already. We require $\delta 1 = 0$.
As a consequence, we have to consider only words of the
type (\ref{3.2}), because (\ref{3.3}) reduces to (\ref{3.2}) for
$a_0 = 1$.\\

\noindent
Finally one defines an operator $\delta$ by the rules
\begin{eqnarray}
\label{3.5}
\delta^2 = 0 \, , & & \\
\delta \left( a_0 \delta a_1 \delta a_2 \cdots \delta a_p \right)
& = & \delta a_0 \delta a_1 \delta a_2 \cdots \delta a_p \nonumber
\end{eqnarray}

\noindent
By linearity, the action of $\delta$ is extended to all elements
of $\Omega A$. We define $\Omega^p A$ to be the linear span of
the words $a_0 \delta a_1 \cdots \delta a_p$, referred to
as ``universal $p$--forms''. Then
\[
\Omega A = \bigoplus^\infty_{p = 0} \,
\Omega^p A \, , \, \, \, \, \, \Omega^0 A \equiv A \, ,
\]

\noindent
is a graded differential algebra with the ``exterior derivative''
$\delta: \Omega^p A \rightarrow \Omega^{p+1} A$.\\

\vspace*{5mm}

\par \noindent
{\Large Defining $\Omega A$ via tensor products}\\

For our purposes it is important to realize that the space
of universal $p$--forms, $\Omega^p A$, can be identified
with a certain subspace of the tensor product
\[
\underbrace{A \otimes A \otimes \cdots \otimes A}_{(p+1)
             \,\, {\rm factors}}
\equiv A^{\otimes (p+1)}
\]
Let us start with a few definitions. The $\otimes_A$--product of
elements from $A^{\otimes (p+1)}$ with elements from
$A^{\otimes (q+1)}$ yields elements in
$A^{\otimes (p + q + 1)}$. It is defined by
\begin{eqnarray}
\left[ a_0 \otimes a_1 \otimes \cdots \otimes a_p \right]
\, \, \otimes_A \, \,
\left[ b_0 \otimes b_1 \otimes \cdots \otimes b_q \right] \nonumber\\
= a_0 \otimes a_1 \otimes \cdots \otimes a_{p-1}
\otimes a_p b_0 \otimes b_1 \otimes \cdots \otimes b_q
\end{eqnarray}

\noindent
where $a_p b_0$ is an ordinary algebra product. Obviously
$\otimes_A$--multiplication is associative. Furthermore,
it is convenient to introduce the multiplication maps
${\rm m}_i \, : \, A^{\otimes (p+1)} \rightarrow A^{\otimes p} \, ,
i = 1, 2, \cdots, p$. They are linear and act as
\begin{eqnarray}
& {\rm m}_i \left[ a_0 \otimes a_1 \otimes \cdots \otimes a_{i-1}
\otimes a_i \otimes \cdots \otimes a_p \right] \\
& = a_0 \otimes a_1 \otimes \cdots \otimes a_{i-1} a_i
\otimes a_{i+1} \otimes \cdots \otimes a_p\nonumber
\end{eqnarray}
In this language, the construction of $\Omega A$ is as follows.
Again we set $\Omega^0 A \equiv A$, and we identify
$ \delta a \in \Omega^1 A$ with the following element of
$A \otimes A$:
\begin{equation}
\label{3.8}
\delta a = 1 \otimes a - a \otimes 1
\end{equation}

\noindent
``Words'' are formed by taking $\otimes_A$--products of $a$'s
and $\delta a$'s:
\begin{equation}
a (\delta b) c (\delta d) \cdots = a \otimes_A
[1 \otimes b - b \otimes 1 ] \otimes_A c \otimes_A
[ 1 \otimes d - d \otimes 1 ] \otimes_A \cdots
\end{equation}

\noindent
A generic element of $\Omega^1 A$ has the structure
\begin{equation}
a \delta b = a \otimes_A [ 1 \otimes b - b \otimes 1 ]
= a \otimes b - ab \otimes 1
\end{equation}

\noindent
Obviously it is in the kernel of the multiplication map ${\rm m}_1$:
${\rm m}_1 (a \delta b) = ab - ab = 0$. More generally one defines
\begin{eqnarray}
\Omega^1 A & = & {\rm Ker} \, \left( {\rm m}_1 \right)  \nonumber \\
\Omega^p A & = & \underbrace{\Omega^1 A \otimes_A \Omega^1 A
\otimes_A \cdots \otimes_A \Omega^1 A}_{p \,\, {\rm factors}}
\end{eqnarray}

\noindent
In this formalism the Leibniz rule (\ref{3.1}) becomes a relation
in $A \otimes A$. It is easy to see that the
identification (\ref{3.8}) is consistent with it:
\begin{eqnarray}
(\delta a) b + a (\delta b) & = & [ 1 \otimes a - a \otimes 1 ]
\otimes_A b + a \otimes_A [ 1 \otimes b - b \otimes 1 ] \nonumber \\
& =  & 1 \otimes ab - ab \otimes 1  \nonumber\\
& = & \delta (ab)
\end{eqnarray}

Let us assume that it is possible to enumerate the
elements of $A$ as $ A= \{ a_n, n \in \Im \}$ where
$\Im$ is some index set whose precise nature we shall not specify
here. Though formal, the following discussion is a useful
preparation for the construction of ``quantum forms''.

We consider two 1--forms $\alpha$ and $\beta$. As $\Omega^1 A$ is
contained in $A \otimes A$, they have expansions of the form
\begin{equation}
\alpha = \sum_{n,m} \alpha_{nm}
a_n \otimes a_m \, , \, \beta = \sum_{k,l}\beta_{k l}
a_k \otimes a_l
\end{equation}

\noindent
Because $\Omega^1 A = {\rm Ker} \, ({\rm m}_1)$
                                the coefficients $\alpha_{nm}$
must be chosen such that
\begin{equation}
\label{3.14}
{\rm m}_1 (\alpha) = \sum_{n,m} \alpha_{nm} a_n a_m = 0
\end{equation}

\noindent
and similarly for $\beta$. The product $\alpha \beta \equiv
\alpha \otimes_A \beta = \Sigma \alpha_{nm} \beta_{kl} \, a_n
\otimes a_m a_k \otimes a_l$ indeed lies in $\Omega^2 A$, because
\begin{equation}
{\rm m}_1 (\alpha \beta) = \sum_{k,l} \beta_{kl}
\left[ \sum_{n,m} \alpha_{nm} a_n a_m \right]
a_k \otimes a_l = 0
\end{equation}

\noindent
vanishes by virtue of (\ref{3.14}), and similarly ${\rm m}_2
                                                       (\alpha \beta)
= 0$. The coefficients of a generic $p$--form $\alpha_p
\in \Omega^p A$ given by the
expansion
\begin{equation}
\label{3.16}
\alpha_p = \sum_{m_0 \cdots m_p}  \, \alpha_{m_0 \cdots
m_p} \, \, a_{m_0} \otimes a_{m_1} \otimes \cdots \otimes
a_{m_p}
\end{equation}

\noindent
are subject to the conditions
 \begin{eqnarray}
 0 = {\rm m}_i \left( \alpha_p \right) = \sum_{m_0 \cdots
 m_p} \alpha_{m_0 \cdots m_p} \, \, a_{m_0}
 \otimes \cdots \otimes a_{m_{i-2}} \\
\otimes a_{m_{i-1}} a_{m_i} \otimes a_{m_{i+1}} \otimes \cdots
\otimes a_{m_p} \nonumber
\end{eqnarray}

\noindent
for $i = 1, \cdots, p$. Therefore (\ref{3.16}) can be rewritten as
\begin{eqnarray}
\alpha_p & = & \sum_{m_0 \cdots m_p}
\alpha_{m_0 \cdots m_p} \, \, a_{m_0} \otimes_A
\left[ 1 \otimes a_{m_1} - a_{m_1} \otimes 1 \right ] \otimes_A
\nonumber \\
& & \, \, \, \, \, \cdots \otimes_A \left[ 1 \otimes a_{m_p} - a_{m_p}
 \otimes 1 \right]\nonumber \\
& & \label{3.18} \\
& = & \sum_{m_0 \cdots m_p} \alpha_{m_0 \cdots m_p}
\, \, a_{m_0} \delta a_{m_1} \delta a_{m_2} \cdots
\delta a_{m_p}  \nonumber
\end{eqnarray}

\noindent
When written in this fashion, each term in the sum is a $p$--form,
and it is now easy to apply the differential $\delta$ to
(\ref{3.18}). It follows directly from the definition
                 (\ref{3.5}) that
$ \delta \alpha_p \in \Omega^{p+1} A$ is represented
by the following element in $A^{\otimes (p+2)}$ :
\begin{eqnarray}
\delta \alpha_p & = & \sum_{m_0 \cdots m_p}
\alpha_{m_0 \cdots m_p} \, \, \delta a_{m_0} \delta a_{m_1} \cdots
\delta a_{m_p} \nonumber \\
& & \label{3.19}\\
& = & \sum_{m_0 \cdots m_p} \alpha_{m_0 \cdots
m_p}
     \bigotimes\limits_{j=0}^p \!\!~_A
\left[ 1 \otimes a_{m_j} - a_{m_j} \otimes 1 \right]\nonumber
\end{eqnarray}

\noindent
Now we need the following relation, which is easily proven by
induction
\begin{eqnarray}
& {\displaystyle \bigotimes\limits^{p}_{j=0}} \!\! ~_A
\left[ 1 \otimes a_{m_j} - a_{m_j} \otimes 1 \right] & \label{3.20}
       \\
& = {\displaystyle \sum\limits^{p+1}_{i=0}} ( -1)^i a_{m_0} \otimes
a_{m_1} \otimes \cdots \otimes a_{m_{i-1}} \otimes 1
\otimes  a_{m_1} \otimes \cdots \otimes a_{m_p}  & \nonumber \\
& \, \, \, \, \,  + \, \, \mbox{irrelevant} & \nonumber
\end{eqnarray}

\noindent
Here ``irrelevant'' stands for terms containing algebra
products $a_{m_{i-1}} a_{m_i}$ inside at least one
factor of the tensor product. These terms vanish upon
contraction with $\alpha_{m_0 \cdots m_p}$. Inserting
(\ref{3.20}) into (\ref{3.19}) we arrive at the final result
\begin{eqnarray}
\delta \alpha_p = \sum^{p+1}_{i=0} (- 1)^i
\sum_{m_0 \cdots m_p} \alpha_{m_0 \cdots m_p}
\, \, a_{m_0} \otimes a_{m_1} \otimes \cdots & & \nonumber  \\
& & \label{3.21} \\
\cdots \otimes a_{m_{i-1}} \otimes 1 \otimes a_{m_i} \otimes
 \cdots \otimes a_{m_p} & & \nonumber
 \end{eqnarray}
In the next section this representation of $\delta \alpha_p$ will
allow us to make contact with the operator $\delta$ defined
on multiscalars.

\renewcommand{\theequation}{4.\arabic{equation}}
\setcounter{equation}{0}

 \section{Quantum Forms on Phase--Space}

In this section we apply the symbol map described in section 1
to $\Omega A$, where $A$ is now taken to be the algebra of
operators acting on the Hilbert space ${\cal H}$. Again
we think of ${\cal H}$ as the space of states of a
certain quantum system with classical phase--space
${\cal M}_{2N}$. In this manner we shall arrive
at a representation of the non--commutative universal
forms which, for $\hbar \rightarrow 0$, connects smoothly
to the standard exterior algebra. In this way, the
non--commutative forms are seen to be a
``deformation'' of the classical ones in the
same sense as the Moyal bracket is a deformation
of the Poisson bracket.

In the introduction we discussed the symbol map
$\, {\rm symb}: A \rightarrow {\rm Fun}\,({\cal M}_{2N}),
\hat{a} \rightarrow {\rm symb} \, (\hat{a}) \equiv a$
which relates operators on ${\cal H}$ to complex functions
over ${\cal M}_{2N}$. We generalize its definition
by including operators $\hat{a}_0 \otimes \hat{a}_1
\otimes \cdots \otimes \hat{a}_p \in A^{\otimes (p+1)}$ which
act on ${\cal H}^{\otimes (p+1)}$. The (linear) map
\[
{\rm symb} \, : A \otimes A \otimes \cdots \otimes A
\rightarrow {\rm Fun}\, ({\cal M}_{2N} \times {\cal M}_{2N}
\times \cdots \times {\cal M}_{2N})
\]
\noindent
represents operators in $A^{\otimes (p + 1)}$ by functions of
$(p + 1)$ arguments:
\begin{eqnarray}
\left[ {\rm symb} \left( \hat{a}_0 \otimes \hat{a}_1 \otimes \cdots
\otimes \hat{a}_p \right) \right]
\left( \phi_0, \phi_1, \cdots, \phi_p \right) \\
 = \left[ {\rm symb} \left( \hat{a}_0 \right) \right]\,
\left(\phi_0 \right)  \,  \left[ {\rm symb} \left( \hat{a}_1
 \right) \right] \left(\phi_1 \right)  \cdots
 \left[ {\rm symb} \left( \hat{a}_p \right) \right]
\left( \phi_p \right)  \nonumber
\end{eqnarray}

\noindent
In this manner we establish a one--to--one correspondence
between the abstract non--commutative differential forms
in $\Omega^p A$ and functions of $p + 1$ arguments. For
instance, if $\alpha_p \in \Omega^p A$ is given by
\begin{equation}
\label{4.2}
\alpha_p = \sum_{m_0 \cdots m_p} \alpha_{m_0 \cdots m_p}
\, \, \hat{a}_{m_0} \otimes \hat{a}_{m_1} \otimes \cdots \otimes
\hat{a}_{m_p}
 \end{equation}

\noindent
then its symbol ${\rm symb} \left( \alpha_p \right) \equiv
F_p$ reads
\begin{equation}
\label{4.3}
F_p \left( \phi_0, \phi_1, \cdots, \phi_p \right)
   = \sum_{m_0 \cdots m_p} \alpha_{m_0 \cdots m_p}
\, \, a_{m_0} \left( \phi_0 \right) a_{m_1} \left( \phi_1  \right)
\cdots a_{m_p} \left( \phi_p \right)
 \end{equation}

\noindent
where we have set $a_m \equiv {\rm symb} \left( \hat{a}_m \right)$.
At the level of symbols, the $\otimes_A$--product becomes
\begin{eqnarray}
\label{4.4}
\left[ {\rm symb} \, \bigg( \left[ \hat{a}_0 \otimes  \right.
\left. \hat{a}_1 \otimes \cdots \otimes \hat{a}_p \right]
\otimes_A  \left[ \hat{b}_0 \otimes \hat{b}_1\right.
 \left. \otimes \cdots \otimes \hat{b}_q  \right] \bigg) \, \right]
 \left( \phi_0, \cdots, \phi_{p+q} \right)  \\
= a_0 \left( \phi_0 \right) a_1 \left( \phi_1 \right)
\cdots a_{p-1} \left( \phi_{p-1} \right)
\left[ a_p \ast  b_0 \right] \left( \phi_p \right)
b_1 \left( \phi_{p+1} \right) \cdots b_q
\left( \phi_{p+q} \right)  \nonumber
\end{eqnarray}

\noindent
with $a_i \equiv {\rm symb} \left( \hat{a}_i \right)$ and
$b_i \equiv {\rm symb} \left( \hat{b}_i \right)$.
The multiplication maps ${\rm m}_i$ act on symbols of the
type (\ref{4.3}) according to
 \begin{eqnarray}
\label{4.5}
 \left( {\rm m}_i F_p \right) \left( \phi_0, \phi_1,
 \cdots, \hat{\phi}_{i-1}, \phi_i, \cdots, \phi_p \right) & & \\
   = \sum_{m_0 \cdots m_p} \alpha_{m_0 \cdots m_p}
                  \, \,    a_{m_0}\left( \phi_0 \right) \cdots
a_{m_{i-2}} \left( \phi_{i-2} \right)
\left[ a_{m_{i-1}} \ast a_{m_i} \right]
\left( \phi_i  \right) & & \nonumber \\
\hspace*{5mm}\cdot a_{m_{i+1}} \left( \phi_{i+1} \right)\cdots a_{m_p}
\left( \phi_p \right)  & & \nonumber
 \end{eqnarray}

\noindent
If $F_p = {\rm symb} \left( \alpha_p \right)$, the condition
$\alpha_p \in \Omega^p A$ turns into
${\rm m}_i F_p = 0, i = 1, \cdots, p$.\\

In the light of these rules we can look at the algebraic
structures of section 3 in either of two ways.
We can identify $A$ with the algebra of operators
$\hat{a}, \hat{b}, \cdots$ (with the notational change
$a \rightarrow \hat{a}$, etc., in section 3 and the
algebra product with the operator multiplication, or we can
identify $A$ with the algebra of symbols equipped
with the $\ast$--product. The most interesting
object to compare in both formulations is the differential
$\delta$. We define its  action on symbols in the obvious
way: $\delta \,{\rm symb} \left( \alpha_p \right) =
{\rm symb} \left( \delta \alpha_p \right)$. For
$\alpha_p$'s of the type (\ref{4.2}), $\delta \alpha_p$ has
been given in eq. (\ref{3.21}). Using (\ref{4.3}) and noting that
\begin{eqnarray}
\label{4.6}
\left( \hat{a}_{m_0} \otimes \hat{a}_{m_1} \otimes
\cdots \otimes \hat{a}_{m_{i-1}} \otimes 1 \otimes
\hat{a}_{m_i} \otimes \cdots \otimes \hat{a}_{m_p} \right)
\left( \phi_0, \cdots, \phi_{p+1} \right) \\
= a_{m_0} \left( \phi_0 \right) \cdots a_{m_{i-1}}
\left( \phi_{i-1} \right) a_{m_i} \left( \phi_{i+1}
\right)  \cdots
a_{m_p} \left( \phi_{p+1} \right) \nonumber
\end{eqnarray}

\noindent
we arrive at an explicit representation of
$
\delta \,: {\rm Fun} \left({\cal M}_{2N}^{(p+1)}\right)
\rightarrow {\rm Fun} \left( {\cal M}_{2N}^{(p+2)} \right),
$
namely
\begin{equation}
\label{4.7}
\left( \delta F_p \right) \left( \phi_0, \phi_1, \cdots,
\phi_{p+1} \right) =  \sum\limits^{p+1}_{i=0} ( -1)^i \ %
F_p \left( \phi_0, \cdots, \phi_{i-1}, \hat{\phi_i},
\phi_{i+1} , \cdots, \phi_{p+1} \right)
\end{equation}

\noindent
In (\ref{4.6}) we assumed that ${\rm symb} (1)$ is the constant
function with value 1, which is true for the Weyl symbol. It is
quite remarkable, that the formula (\ref{4.7}) coincides exactly
with eq. (\ref{2.2}), which was at the heart of our multiscalar
approach to the classical exterior algebra. There remains
an important difference however. The forms
$F_p \in \bigwedge_{\rm MS}^p({\cal M}_{2N})$ studied in
section 2 were supposed to vanish when two adjacent arguments
are equal. The symbols $F_p = {\rm symb} \left( \alpha_p \right)$,
instead, obey a deformed version of this condition, namely
${\rm m}_i F_p = 0$. In fact, in the classical (i.e. commutative)
limit $\hbar \rightarrow 0$, the star--product becomes the
ordinary point--wise product of functions, and (\ref{4.5})
yields
\begin{equation}
\left( {\rm m}_i F_p \right) \left( \phi_0, \cdots, \hat{\phi}_{i-1},
\phi_i, \cdots, \phi_p \right)  =  F_p \left( \phi_0, \cdots,
\phi_{i-2}, \phi_i, \phi_i, \cdots, \phi_p \right)
 +O(\hbar)
\end{equation}

\noindent
so that the conditions are the same in both cases. Similary, the
product (\ref{4.4}) reduces to the product (\ref{2.7})
                                           for $\hbar \rightarrow
0$. Therefore we may conclude that in the classical limit
$\Omega^p A$ and $\bigwedge^p_{\rm MS}({\cal M}_{2N})$ are
equivalent. Symbolically,
$$
\lim_{\hbar \to 0} \, {\rm symb} \, \left( \Omega^p A
\right) = \bigwedge\nolimits^p_{\rm MS} ({\cal M}_{2N})
$$

As shown in section 2, the ordinary exterior algebra
$\bigwedge({\cal M}_{2N})$ is obtained from $\bigwedge_{\rm MS}
({\cal M}_{2N})$ by an appropriate coincidence limit. The
interesting question which we will address in section 5,
is what happens if we perform this coincidence limit
for $\hbar \neq 0$.\\

\vspace*{5mm}

\par \noindent
{\large The quantum--deformed Lie derivative}\\

Let us fix a certain $\alpha_p \in \Omega^p A$ with an
expansion of the form (\ref{4.2}). It is an operator on the
$(p+1)^{st}$ tensor power of the Hilbert space ${\cal H}$,
${\cal H}^{\otimes (p+1)}$. Let us now perform the same
unitary transformation on all factors of
${\cal H}^{\otimes (p+1)}$. We assume that the
$\hat{a}_m$'s in (\ref{4.2}) are transformed according to
\begin{equation}
\hat{a}_{m_j} (t) = \exp \left( - \frac{i}{\hbar}
\hat{H} t \right) \hat{a}_{m_j} \exp
\left( \frac{i}{\hbar} \hat{H} t \right)
\end{equation}

\noindent
where $\hat{H}$ is a hermitian generator and $t$ is a
real parameter. Thus the family of operators
$\hat{a}_{m_j} (t)$ obeys the von Neumann--type
equation
\begin{equation}
i \hbar \, \partial_t \, \hat{a}_{m_j} (t) =
\left[ H, \hat{a}_{m_j} (t) \right]
\end{equation}

\noindent
and the evolution of $\alpha_p$ as a whole is governed by
\begin{eqnarray}
i \hbar \, \partial_t \, \alpha_p (t) &  = & \sum\limits^p_{j=0}
\sum\limits_{m_0 \cdots m_p} \alpha_{m_0 \cdots m_p}\, \,
                             \hat{a}_{m_0} (t) \otimes \cdots
                                              \nonumber \\
& & \label{4.10}\\
& \cdots & \otimes \left[ \hat{H}, \hat{a}_{m_j} (t) \right]
\otimes \cdots \otimes \hat{a}_{m_p} (t)\nonumber
\end{eqnarray}

\noindent
If we apply the Weyl symbol map to both sides of this
equation and use (\ref{1.9}) and (\ref{1.10}) with
             (\ref{1.11}) we arrive at
\begin{equation}
- \partial_t F_p \left( \phi_0, \cdots, \phi_p; t \right) =
{\cal L}^{\hbar}_p \left[ H \right] F_p \left( \phi_0,
\cdots, \phi_p; t \right)
\end{equation}

\noindent
with the ``quantum deformed Lie derivative''
\begin{equation}
{\cal L}^{\hbar}_p \left[ H \right] =
- \sum\limits^p_{i=0} \, \frac{2}{\hbar} \, H
\left( \phi_i \right) \sin
\left[ \frac{\hbar}{2} \, \stackrel{\leftarrow}{\partial}_a^{(i)}
                                                   \omega^{ab}\,
\stackrel{\rightarrow}{\partial}_b^{(i)} \right]
\end{equation}

\noindent
and $F_p \equiv {\rm symb} \left( \alpha_p \right)$,
$H \equiv {\rm symb} \left( \hat{H} \right)$.

In the limit $\hbar \rightarrow 0$, ${\cal L}_p^\hbar$
reduces to the classical Lie derivative (\ref{2.12}) appropriate for
multiscalars. This suggests the interpretation of the symbols
${\rm symb} (\alpha), \alpha \in \Omega A$, as quantum
deformed multiscalars. When a classical multiscalar
is subject to a canonical transformation, the hamiltonian
vector field $-V_H = \omega^{ba} \partial_a H \partial_b$
acts on any of its arguments. In the non--commutative case,
$V_H$ is replaced by its Moyal analogue, eq. (\ref{1.11}).
Like the classical one, the quantum Lie derivative
commutes with the differential $\delta$ :
\begin{equation}
\delta_p \, {\cal L}_p^\hbar = {\cal L}_{p+1}^\hbar \, \delta_p
\end{equation}

\noindent
As a consequence of the algebra (\ref{1.12}) for the deformed
hamiltonian vector fields, the deformed Lie derivatives form
a closed algebra as well:
\begin{equation}
\left[ {\cal L}_p^\hbar \left[ H_1 \right],
{\cal L}_p^\hbar \left[ H_2 \right] \right] =
- {\cal L}_p \left[ \{H_1, H_2\}_{\rm mb} \right]
\end{equation}

\noindent
This is the algebra of {\it quantum} canonical
transformations, which is closely related to the
$W_{\infty}$ algebra \cite{fz, winf}.
In ref. \cite{winf} it was shown that, under the symbol map,
the algebra of infinitesimal unitary transformations on Hilbert
space translates to an algebra of the above type.
Choosing a basis on the space of all generating functions $H$
one arrives at the more familiar forms of the $W_\infty$-algebra
\cite{fz} in which the structure constants are given by the
Moyal brackets of the basis functions.

\renewcommand{\theequation}{5.\arabic{equation}}
\setcounter{equation}{0}

 \section{Coincidence Limit in the Non-Commuta- \newline tive Case}

In this section we investigate the coincidence limit of the Moyal
multiscalars $F_p = {\rm symb} \left( \alpha_p \right)$.
Differences relative to the classical discussion in section 2
will occur because the pointwise product of functions
on ${\cal M}_{2N}$ is now replaced by the star product. The
impact of this deformation on the properties of
differential forms is best illustrated by means of a few
examples.

First we consider a 1--form $\alpha_1 \in \Omega^1 A$
represented by
\begin{equation}
\alpha_1 = \hat{a}_0 \delta   \hat{a}_1 =
\hat{a}_0 \otimes \hat{a}_1 - \hat{a}_0 \hat{a}_1 \otimes 1
\end{equation}
\noindent
Writing as usual $a_i = {\rm symb} \, \left( \hat{a}_i \right)$,
the symbol of $\alpha_1$ reads
\begin{eqnarray}
\label{5.2}
F_1 \left( \phi_0, \phi_1 \right) & = &
a_0 \left( \phi_0 \right) a_1 \left( \phi_1 \right) -
\left( a_0 \ast a_1 \right) \left( \phi_0 \right) \\
& = & F_1^{\rm class} \left( \phi_0, \phi_1 \right) +
\left( a_0 a_1 - a_0 \ast a_1 \right) \left( \phi_0 \right)\nonumber
\end{eqnarray}
\noindent
with
\begin{equation}
F_1^{\rm class} \left( \phi_0, \phi_1 \right) \equiv a_0
\left( \phi_0 \right) \left[ \right. a_1 \left( \phi_1 \right) -
\left. a_1 \left( \phi_0 \right) \right]
\end{equation}

\noindent
In the second line of (\ref{5.2}) we decomposed $F_1$ in
the classical part $F_1^{\rm class}$ and a quantum
correction which vanishes for $\hbar \rightarrow 0$.
Obviously $F_1^{\rm class}$ is of the type (\ref{2.29}), and
eq. (\ref{2.30}) tells us that in the coincidence limit
$F_1^{\rm class} \left( \phi, \phi + \eta \right) =
\eta^b a_0 \partial_b a_1 + 0 \left( \eta^2 \right)$. Therefore
the full $F_1$ yields
\begin{equation}
\label{5.4}
F_1 \left( \phi, \phi + \eta \right) = \eta^b a_0 \left( \phi \right)
 \partial_b a_1 \left( \phi \right) + \left( a_0 a_1 -
 a_0 \ast a_1 \right) \left( \phi \right) +
 O \left( \eta^2 \right)
 \end{equation}
Contrary to the classical multiscalars in $\bigwedge^p_{\rm MS}
({\cal M}_{2N})$, the Moyal multiscalars $F_p$ are not simply
proportional to $ \eta_1^{a_1} \cdots \eta_p^{a_p}$ in
the coincidence limit: there are also terms with
$ \eta_1^{a_1} \cdots \eta_l^{a_l}, l < p$. In (\ref{5.4}) this
general rule is illustrated by the $\eta$--independent
quantum correction $a_0 a_1 - a_0 \ast a_1$.
To be more explicit, let us choose $\hat{a}_1 = \hat{\phi}^b$ for
some fixed index $b$. Hence $\hat{a}_1$ is one of the
canonical operators $\left( \hat{p}^1, \cdots,
\hat{p}^N; \hat{q}^1, \cdots, \hat{q}^N \right)$, and
the associated Weyl symbol is simply $a_1 \left( \phi \right) =
\phi^b$. In this case the series expansion for the
star product terminates after the second term:
\begin{equation}
\left[ {\rm symb} \left( \hat{a}_0 \delta \hat{\phi}^b \right) \right]
 \left( \phi, \phi + \eta \right) = a_0 \left( \phi \right)
 \eta^b + \frac{1}{2} i \hbar \omega^{bc} \partial_c a_0
 \left( \phi \right)
 \end{equation}

\noindent
The most unusual property of the non--commutative 1--form
$\hat{a}_0 \delta \hat{\phi}^b$ is that its symbol does not
vanish even at coinciding points:
\begin{equation}
\label{5.6}
\left[ {\rm symb} \left( \hat{a}_0 \delta \hat{\phi}^b \right) \right]
\left( \phi, \phi \right) = \frac{1}{2} i \hbar \omega^{bc}
\partial_c a_0 \left( \phi \right)
\end{equation}

\noindent
Note that the RHS of (\ref{5.6}) is purely imaginary and that it
is proportional to be $b$--component of the hamiltonian
vector field generated by $a_0$. The nonvanishing RHS
of (\ref{5.6}) is a typical quantum effect. It seems to
contradict our ``classical'' intuition about the
meaning of a differential $d \phi^b$. Loosely speaking,
given two ``nearby'' points $\phi_0$ and $\phi_1$,
we would like to visualize $d \phi^b$ as the
``displacement'' $\phi_1^b - \phi_0^b \equiv \eta^b$.
Consequently we expect that, in an appropriate sense,
$d \phi^b = 0$ if $\phi_0 = \phi_1$. {\it Eq.(\ref{5.6}) shows
that this is not necessarily the case for quantum
1--forms.} Though it is true that ${\rm symb} \, (\delta \hat{\phi}^b
)$ vanishes in
the coincidence limit, this is not the case
anymore as soon as we multiply $\delta \hat{\phi}^b$
by some nontrivial operator $\hat{a}_0$.

Next we look at a non--commutative 2--form $\alpha_2 =
\delta   \hat{a}_0 \delta \hat{a}_1 \in \Omega^2 A$.
Its tensor product representation
\begin{eqnarray}
\label{5.7}
\alpha_2 & = & \left[ 1 \otimes \hat{a}_0 - \hat{a}_0 \otimes 1 \right]
\otimes_A \left[ 1 \otimes \hat{a}_1 - \hat{a}_1 \otimes 1 \right] \\
& = & 1 \otimes \hat{a}_0 \otimes \hat{a}_1 -
1 \otimes \hat{a}_0 \hat{a}_1 \otimes 1 - \hat{a}_0 \otimes
1 \otimes \hat{a}_1 + \hat{a}_0 \otimes \hat{a}_1 \otimes
1 \nonumber
\end{eqnarray}

\noindent
translates into the symbol
\begin{equation}
\label{5.8}
F_2 \left( \phi_0, \phi_1, \phi_2 \right) = F_2^{\rm class}
\left( \phi_0, \phi_1, \phi_2 \right) + \left( a_0 a_1 -
a_0 \ast a_1 \right) \left( \phi_1 \right)
\end{equation}

\noindent
with
\begin{equation}
F_2^{\rm class} \left( \phi_0, \phi_1, \phi_2 \right) =
\left[ a_0 \left( \phi_1 \right) - a_0 \left( \phi_0 \right) \right]
\left[ a_1 \left( \phi_2 \right) - a_1 \left( \phi_1 \right) \right]
\end{equation}

\noindent
The non--classical piece in (\ref{5.8}) is due to the operator
product $\hat{a}_0 \hat{a}_1$ in the second line of (\ref{5.7}).
Using (\ref{2.17}) and (\ref{2.30}) for the expansion
of $F_2^{\rm class}$, we obtain
 \begin{eqnarray}
\label{5.10}
 F_2 \left( \phi, \phi + \eta_1, \phi + \eta_2 \right)
 & = & \left( a_0 a_1 - a_0 \ast a_1 \right)
 \left( \phi \right) + \eta_1^b \partial_b
 \left( a_0 a_1 - a_0 \ast a_1 \right) \left( \phi \right)
 \nonumber                                   \\
& & \\
 & & + \eta_1^b \eta_2^c\  \partial_b a_0
 \left( \phi \right) \partial_c a_1 \left( \phi \right) +
 O \left( \eta_1^2, \eta_2^2 \right) \nonumber
 \end{eqnarray}

\noindent
Apart from the classical term proportional to
$\eta_1 \eta_2$, we find a term linear in $\eta$ and
a constant piece which survives the limit $\eta_1, \eta_2
\rightarrow 0$. Eq.~(\ref{5.10}) becomes particularly transparent
for the choice $\hat{a}_0 = \hat{\phi}^a,
\hat{a}_1 = \hat{\phi}^b$ with fixed indices $a$ and $b$:
\begin{eqnarray}
\label{5.11}
\left[ {\rm symb} \left( \delta \hat{\phi}^a \delta
\hat{\phi}^b \right)  \right] \left( \phi, \phi + \eta_1,
\phi + \eta_2 \right) \\
= - \frac{1}{2} \, i \hbar \omega^{ab} + \eta_1^a \eta_2^b +
O \left( \eta_1^2, \eta_2^2 \right) \nonumber
\end{eqnarray}

\noindent
Upon antisymmetrization, the term $\eta_1^a \eta_2^b$ gives
rise to the classical analogue of $\delta \hat{\phi}^a
\delta \hat{\phi}^b$, namely $d \phi^a \wedge d \phi^b$.
In a symbolic notation the modified wedge product reads
\begin{equation}
\label{5.12}
\delta \hat{\phi}^a \wedge \delta \hat{\phi}^b \equiv
\delta \hat{\phi}^a \delta \hat{\phi}^b - \delta \hat{\phi}^b
\delta \hat{\phi}^a \sim - i \hbar \omega^{ab} +
d \phi^a \wedge d \phi^b
\end{equation}

\noindent
A natural candidate for a ``quantum deformed symplectic 2--form''
is
\begin{equation}
\label{5.13}
\hat{\omega}_q = \frac{1}{2} \, \omega_{ab} \, \delta \hat{\phi}^a
\wedge \delta \hat{\phi}^b
\end{equation}
\noindent
so that $\hat{\omega}_q \sim \omega_{\rm class} + i N \hbar$.
Using (\ref{5.7}) and $\omega_{ab} \hat{\phi}^a \hat{\phi}^b =
- i \hbar N$, which follows from the canonical commutation
relations, $\hat{\omega}_q$~can be written as
\begin{eqnarray}
\hat{\omega}_q & = & \omega_{ab}
\left[ 1 \otimes \hat{\phi}^a \otimes \hat{\phi}^b - \right.
                                                     \hat{\phi}^a
\otimes 1 \otimes \hat{\phi}^b + \hat{\phi}^a \otimes
\hat{\phi}^b \otimes 1 \left. \right] \nonumber \\
& & + i \hbar N \left( 1 \otimes 1 \otimes 1 \right)
\end{eqnarray}

\noindent
Even without invoking the coincidence limit its symbol
$\omega_q$ has a rather transparent structure:
\begin{equation}
\label{5.15}
\omega_q \left( \phi_0, \phi_1, \phi_2 \right) =
\omega_{ab}
\left[ \phi_1^a \phi_2^b - \phi_0^a \phi_2^b + \phi_0^a \phi_1^b \right]
+ i N \hbar
\end{equation}

\noindent
For $\hbar = 0$ the function $\omega_q \left( \phi_0, \phi_1,
\phi_2 \right)$ is precisely (twice) the symplectic area of the
triangle with vertices \footnote{Recall that we are using
Darboux canonical coordinates on ${\cal M}_{2N} = {\bf R}^{2N}$.}
$\phi_0, \phi_1$ and $\phi_2$. Clearly this area vanishes
when the edges of the triangle shrink and its vertices
merge in a single point. The situation is dramatically
different for $\hbar \neq 0$. If we use
(\ref{5.15}) to define a symplectic area also in the
quantum case, we find that this ``area'' is nonzero even
for degenerate triangles with coincident vertices:
\begin{equation}
\label{5.16}
\omega_q \left( \phi, \phi, \phi \right) = i N \hbar
\end{equation}
Accepting this definition of a ``quantum area'' we see that in
the non--commutative case there is always a
mininal symplectic area of order $\hbar$. It is tempting to
relate this minimum area to the well-known statement that
``it is impossible to localize a quantum state in a
phase--space volume smaller than $\left( 2 \pi \hbar \right)^N$''.
In fact, for $N = 1$, $\hat{\omega}_q$ is the
volume form, and from the derivation of (\ref{5.16}) it is clear
that both phenomena have the same origin, namely the
non--commutative nature of $\hat{p}$ and $\hat{q}$ or,
equivalently, of the
star--product. However, as the quantum mechanical term
in the 2--form $\delta \hat{p} \wedge \delta \hat{q} \sim
i \hbar + d p \wedge dq$ is purely imaginary, the naive
picture of phase--space being partitioned into
``cells'' of volume $\hbar$ cannot be taken too literally.

\renewcommand{\theequation}{6.\arabic{equation}}
\setcounter{equation}{0}

 \section{Discussion and Conclusion}

In this paper we obtained a representation of the
universal algebra of non--com\-mu\-ta\-tive differential forms
on the phase--space of an arbitrary quantum system
by applying the Weyl--Wigner symbol map to the operatorial
construction. The resulting quantum $p$--forms are
multiscalar functions of $p+1$ phase--space variables. Their
coincidence limit yields a deformation of the
classical exterior algebra. At the operatorial level there
exists a natural definition of an exterior derivative
and of a Lie derivative for non--commutative forms.
Their image under the symbol map leads to the
corresponding operations acting on multiscalars. In the
coincidence limit these derivations yield a deformation
of the classical exterior derivative and of the Lie
derivative, to which they reduce for $\hbar \rightarrow
0$. For $\hbar > 0$ the Lie derivative is a generalization
of the Moyal bracket. Lie derivatives
belonging to different hamiltonian vector fields form a
closed algebra of the $W_{\infty}$--type.

            A priori,
the quantum $p$--forms, seen as elements of
$A \otimes A \otimes \cdots \otimes A$, seem to be of
a very different nature than the classical  tensor
fields. In our approach both of them can be represented
within the same setting, and the non--commutative
case can be studied as a smooth deformation
of the classical one.

The non--commutative exterior algebra developped in this
paper contains the standard phase--space formulation of
quantum mechanics as its zero--form sector. The same is also
true for an earlier, different model \cite{gr} of a non--commutative
symplectic geometry. However, in \cite{gr} a
slightly ad hoc definition of a quantum differential
form was used, which led to the unusual feature that
the algebra of Lie derivatives closed only on a space
larger than that of the classical Hamiltonians. The
non--commutative exterior algebra obtained in the present paper is also
different from the ones studied in refs. \cite{dbsym} and
\cite{cart}. Also Segal's ``quantized deRham complex"
\cite{seg} is based upon a different notion of a quantum
differential form.

Before closing we would like to emphasize that the
non--commutative forms studied here are interesting
objects also from a dynamical point of view. In physical
terms the transition from the algebra $A$ to $\Omega^p A$
means that we go over from a one--particle theory living
on the Hilbert space ${\cal H}$ to a $(p+1)$--particle
theory which lives on the tensor product
${\cal H} \otimes {\cal H} \otimes \cdots \otimes {\cal H}$
with $p+1$ factors. The dynamics for all $p+1$ particles
is exactly the same, and no explicit interactions
between the particles are introduced, see eq.~(\ref{4.10}).
Nevertheless, by the very definition of the non--commutative
tensor product, particles belonging to different factors
of ${\cal H} \otimes \cdots \otimes {\cal H}$ actually
``know'' about each other. To make this more explicit
we first consider Hamilton's equation
$\partial_t \phi^a (t) = h^a (\phi (t))$ and linearize
it about a given solution $\phi^a (t)$. This leads
to Jacobi's equation $\partial_t \Delta^a (t) =
\partial_b h^a (\phi) \Delta^b (t)$ which tells us
how the ``displacement'' $\Delta^a$ between the classical
trajectories $\phi^a (t)$ and $\phi^a (t) + \Delta^a (t)$
evolves with time. The ``Jacobi field'' $\Delta^a (t)$
defines a family of classical $1$--forms along
$\phi^a (t)$. The corresponding non--commutative
construction is as follows. From the canonical
operators $\hat{\phi}^a (t)$ in the Heisenberg
picture we define the quantum  $1$--form
\begin{equation}
\label{6.1}
\delta \hat{\phi}^a (t) = 1 \otimes \hat{\phi}^a (t)
- \hat{\phi}^a (t) \otimes 1
\end{equation}

\noindent
whose symbol is $\phi_1^a (t) - \phi^a_0 (t) \equiv
\Delta^a_q (t)$. Under the symbol map the Heisenberg
 equation for $\hat{\phi}^a (t)$ becomes
 Hamilton's equation for the symbols
 $\phi^a (t)  \equiv {\rm symb} \, \left(
 \hat{\phi}^a (t) \right)$. Therefore
\begin{eqnarray}
\partial_t \Delta^a_q (t) & = & h^a \left( \phi_1 (t) \right)
- h^a \left( \phi_0 (t) \right) \\
& = & \partial_b h^a \left( \phi_0 \right)
\Delta^b_q (t) + 0 \left(\Delta^2_q \right) \nonumber
\end{eqnarray}

\noindent
and $\Delta^a_q$ can be identified with the classical
Jacobi field $\Delta^a$ in the coincidence limit
$\phi_1 \rightarrow \phi_0$. This simple observation
illustrates why a non--commutative 1--form
is naturally related to a quantum system on the doubled
   Hilbert space ${\cal H} \otimes {\cal H}$ : in
order to define $\delta \hat{\phi}^a (t)$ via some
kind of ``displacement'' we have to know the position
of {\it two} particles at each instant of time.

The 1--form (\ref{6.1}) is still an essentially classical
object. This is different for the higher $p$--forms
\begin{equation}
\label{6.3}
\delta \hat{\phi}^{a_1} (t) \wedge \delta \hat{\phi}^{a_2}
 (t) \wedge \cdots \wedge \delta \hat{\phi}^{a_p} (t)
 \end{equation}

\noindent
which represent $p$--volumes transported along the
hamiltonian flow. They can be visualized as parallelepipeds
with vertices $(\phi_0, \phi_1, \cdots, \phi_p )$
which should be thought of as the arguments of the
symbol of (\ref{6.3}). Because of the non--commutative
tensor product involved in (\ref{6.3}), the coincidence limit
of this symbol contains the typical $\hbar \omega^{ab}$--terms
studied in section 5. For instance, if we have three nearby
trajectories $\phi_0 (t), \phi_1 (t)$ and $\phi_2 (t)$ the
above construction leads to a two--dimensional area which
is dragged along the hamiltonian flow, and in the symbolic
notation of eq.~(\ref{5.12}) we obtain the time--dependent
area element
\begin{equation}
\delta \hat{\phi}^a (t) \wedge \delta \hat{\phi}^b (t) \sim
i \hbar \omega^{ab} + d \phi^a (t) \wedge d \phi^b (t)
\end{equation}

\noindent
Here we see very clearly that, though individually each
trajectory is governed by the standard one--particle
dynamics, their tensorization leads to the quantum
correction $i \hbar \omega^{ab}$ as a collective effect.
In physical terms it expresses the fact that the
(imaginary) area of the parallelogram with
vertices $\phi_0 (t), \phi_1 (t)$ and $\phi_2 (t)$
is bounded below by $\hbar$.

In a classical context, time--dependent $p$--volume elements
of the type (\ref{6.3}) play an important r$\hat{\rm o}$le in the study
of the chaotic behavior of a system \cite{ll}. The exponential
growth rate of (\ref{6.3}) defines the $p-th$ Lyapunov exponent.
If it is non--zero, the classical evolution of the
system shows a strong form of stochasticity. It is very
natural to ask whether the non--commutative
$p$--forms play a similar r$\hat{\rm o}$le for
``quantum chaos''. We shall come back to this point elsewhere.

  The work contained in this paper can be generalized and
  extended in various directions. Here we assumed, for
instance, that the phase space under consideration is
${\bf R}^{2N}$. Using recent results by Fedosov \cite{fed} it
seems possible to generalize the construction    to
arbitrary symplectic manifolds. Another point which has
to be explored further is whether there are natural
candidates for vectors dual to the non--commutative
forms discussed here. (Contrary to the approach of
Dubois--Violette \cite{dbsym} this is not
self--evident if one starts from \cite{co}.)
As for applications, it is clear that in order
to better understand the dynamical and the geometrical
meaning of the non--commutative forms concrete
examples should be worked out. The important question
is whether, given a conventional quantum system on
${\cal H}$, the dynamics of the higher differential
forms on ${\cal H} \otimes \cdots \otimes {\cal H}$ gives
us useful information about the original system. In
the classical theory the Lyapunov exponents are a
typical example where this actually happens. The
higher $p$--form sectors encode information about
the dynamics of the zero--form sector in a very
transparent way.

Investigations along these lines are not necessarily
restricted to systems where ${\cal M}_{2N}$ is the
true phase--space. For example, one could also
consider planar fermion systems in strong magnetic
fields where the configuration space is effectively
turned into a phase--space \cite{mag}, or the model of
Doplicher et al. \cite{freden} in which quantum
gravity effects induce a kind of symplectic structure
in space--time.

\subsection*{Acknowledgements}

I would like to thank E.~Gozzi for many stimulating
discussions and the Department of Theoretical Physics,
University of Trieste, for its hospitality while
this work was in progress. I am also grateful to
F.~Benatti, M.~Dubois--Violette and N.~Papadopoulos
for several helpful conversations. This work was supported
in part by INFN and by a NATO travelling grant.

 \newpage

\renewcommand{\theequation}{A.\arabic{equation}}
\setcounter{equation}{0}

\subsection*{Appendix A}

In this appendix we derive eq.~(\ref{2.20}) of section 2.
We shall exploit the identity
\begin{equation}
 \label{A.1}
\partial^{(l_1)}_{a_1} \cdots \partial^{(l_M)}
_{a_M} \left[ \partial^{(i)}_b + \partial^{(i+1)}_b \right]
       F_p \, (\phi, \cdots, \phi)= 0
\end{equation}

\noindent
which holds for any $F_p \in \bigwedge^p_{\rm MS}$ provided
$l_k \neq i, i+1$ for all $k = 1, \cdots, M$. Eq.~(\ref{A.1}) follows
simply from the fact that the function defined by
\[
G(\varphi)  = F_p \left( \phi_0, \phi_1, \cdots, \phi_{i-1},
\varphi, \varphi, \phi_{i+2}, \cdots, \phi_p \right)
\]

\noindent
vanishes identically in $\varphi$, and therefore also all its partial
derivatives are zero:
\[
\frac{\partial}{\partial \varphi^b} G(\varphi) =
\left[ \partial^{(i)}_b + \partial^{(i+1)}_b \right]
F \left(\phi_0, \cdots, \varphi, \varphi, \cdots, \phi_p \right)
= 0
\]

\noindent
Taking derivatives with respect to the ``parameters''
$\left( \phi_0, \phi_1, \cdots,
\hat{\phi}_i,\hat{\phi}_{i+1},\cdots, \phi_p \right)$
    we still get zero on the RHS, which proves (\ref{A.1}).

In order to prove (\ref{2.20}) we rewrite (\ref{2.18}) according to
\begin{eqnarray}
\label{A.2}
& F_p  \left( \phi, \phi + \eta_1, \cdots, \phi + \eta_p \right)
&   \nonumber \\
& = {\displaystyle \prod\limits^p_{i=1}}
                           \exp \left[ \eta_i^a \partial^{(i)}_a \right]
F_p \left( \phi, \phi, \cdots, \phi \right) &  \nonumber \\
& = {\displaystyle \prod\limits^p_{i=1}}
                         \left[ 1 + \eta^a_i \partial^{(i)}_a \right]
F_p \left(\phi, \phi, \cdots, \phi \right) + O \left( \eta^2_i \right)
&
\end{eqnarray}

\noindent
In the expansion of the exponential we omitted terms with two
or more equal factors of $\eta_i$. If we perform the product
in the last line of (\ref{A.2}) we seem to obtain tensors of any rank
between zero ($p$ factors of ``1'') and $p$
($p$ factors of $\eta^a_i \partial_a^{(i)}$). However, for
multiscalars $F_p \in \bigwedge^p_{\rm MS}$ we shall
now show that
\begin{equation}
\label{A.3}
\partial^{(l_1)}_{a_1} \cdots \partial^{(l_r)}_{a_r}
F_p \left( \phi, \phi, \cdots, \phi \right) = 0 \hspace*{1cm} \forall
r = 1, \cdots, p-1
\end{equation}

\noindent
with all $l_i$'s different. This means that only the $p$--form piece
in (\ref{A.2}) is non--zero.

The function $F_p$ in eq.~(\ref{A.3}) has $p+1$ arguments which
are acted upon by at most $p-1$ derivatives. All of them
act on different arguments of $F_p$. Hence $F_p$ has
$(p+1) - r$ arguments which are not differentiated.
If two of these arguments are next to each other, we
obtain zero immediately. In order to illustrate the
situation when no undifferentiated arguments are
next to each other, let us consider the most
``dangerous'' case $r = p-1$ with only two
undifferentiated arguments, $\phi_0$ and $\phi_p$, say.
By repeatedly applying (\ref{A.1}) we can write
\begin{eqnarray}
\label{A.4}
\hspace*{2.2cm} \partial^{(1)}_{a_1} \partial^{(2)}_{a_2} \cdots
\partial^{(p-2)}_{a_{p-2}} \partial^{(p-1)}_{a_{p-1}}
F_p \left( \phi, \phi, \cdots, \phi \right) \hspace*{15mm}
                                            & & \nonumber \\
& & \nonumber  \\
= - \partial^{(1)}_{a_1} \partial^{(2)}_{a_2} \cdots
\partial^{(p-3)}_{a_{p-3}} \partial^{(p-2)}_{a_{p-2}}
\partial^{(p)}_{a_{p-1}}
F_p \left( \phi, \phi, \cdots, \phi \right) & & \nonumber \\
& & \nonumber  \\
= + \partial^{(1)}_{a_1} \partial^{(2)}_{a_2} \cdots
\partial^{(p-3)}_{a_{p-3}} \partial^{(p-1)}_{a_{p-2}}
\partial^{(p)}_{a_{p-1}}
F_p \left( \phi, \phi, \cdots, \phi \right)
                                            & & \nonumber \\
& & \nonumber  \\
= \mp \cdots \hspace*{6.9cm} & &
\end{eqnarray}

\noindent
The argument which is not differentiated moves from the right
to the left. This process can be continued until it reaches
the position adjacent to $\phi_0$. At this point
(\ref{A.4}) is seen to vanish. In the same way one
concludes that (\ref{A.3}) is true in general.

\renewcommand{\theequation}{B.\arabic{equation}}
\setcounter{equation}{0}

\subsection*{Appendix B}

In this appendix we prove eqs.~(\ref{2.23}) and (\ref{2.24}) of
section 2.
Let us start by showing that
\begin{equation}
\label{B.1}
{\rm Cl} \, \left( \delta F_p \right) = d \, {\rm Cl}
\left( F_p \right)
\end{equation}

\noindent
We have to extract the monomial with $p$ different factors
of $\eta$ from
\[
D \left( \eta_1, \cdots, \eta_{p+1} \right) \equiv
\left( \delta F_p \right) \left( \phi, \phi +
\eta_1, \cdots, \phi + \eta_{p+1}   \right)
\]

\noindent
 Applying the definition of $\delta$, eq.~(\ref{2.2}), yields
 $ D = D_1 + D_2$ with
 \begin{eqnarray}
\label{B.2}
 D_1 \left( \eta_1, \cdots, \eta_{p+1} \right) & \equiv&
 F_p \left( \phi + \eta_1, \phi + \eta_2, \cdots,
 \phi + \eta_{p+1} \right)            \\
D_2 \left( \eta_1, \cdots, \eta_{p+1} \right) & \equiv&
\sum\limits^{p+1}_{i=1} \left( -1 \right)^i
F_p \left( \phi, \phi + \eta_1, \cdots,
\widehat{\phi + \eta_i}, \cdots,
                             \phi + \eta_{p+1} \right)  \nonumber
\end{eqnarray}

\noindent
Here we have separated the term with $i = 0$. Ignoring
irrelevant terms with higher powers of $\eta_i$, its
contribution reads
\begin{eqnarray}
\label{B.3}
D_1 \left( \eta_1, \cdots, \eta_{p+1} \right) & = & \exp
\left[ \sum\limits^{p+1}_{i=1} \left( \eta_i \right.
\left. \partial^{(i-1)} \right)  \right]
          F_p (\phi, \phi, \cdots, \phi ) \\
& = & \prod\limits^{p+1}_{i=1} \left[ 1 + \left( \eta_i \right.
\left. \partial^{(i-1)} \right) \right] F_p (\phi, \phi, \cdots, \phi)
+ \cdots \nonumber
\end{eqnarray}

\noindent
where $\left( \eta_i \, \partial^{(i-1)} \right) \equiv
\eta^a_i \partial^{(i-1)}_a$. By eq.~{(\ref{A.3}) of
appendix A the coincidence limit of the monomials
$\partial \partial \cdots \partial \, F_p$ is zero if there are not at
least $p$ (different) derivatives acting
on $F_p$. This means that when we expand the
product in the last line of (\ref{B.3}) we have to keep only the
terms with $p+1$ derivatives and with $p$ derivatives:
\begin{eqnarray}
\label{B.4}
 D_1 & = & D^p_1 + D^{p-1}_1  \\
D^p_1 & \equiv & \left( \eta_1 \partial^{(0)} \right)
\left( \eta_2 \partial^{(1)} \right) \cdots
\left( \eta_{p+1} \partial^{(p)} \right) \,
F_p \, (\phi, \phi, \cdots, \phi )  \nonumber \\
D^{p-1}_1 & \equiv & \sum\limits^{p+1}_{i=1}
\left( \eta_1 \partial^{(0)} \right)
\left( \eta_2 \partial^{(1)} \right) \cdots
\left( \eta_{i-1} \partial^{(i-2)} \right)
\left( \eta_{i+1} \partial^{(i)} \right) \cdots  \nonumber \\
& &  \hspace*{35mm} \cdots \left( \eta_{p+1} \partial^{(p)} \right)
 F_p \, (\phi, \cdots, \phi)  \nonumber
 \end{eqnarray}

\noindent
The relevant piece in $D_2$ of (\ref{B.2}) is
 \begin{eqnarray}
\label{B.5}
 D_2 \left( \eta_1, \cdots, \eta_p \right) =
  \sum\limits^{p+1}_{i=1} (-1)^i
 \left( \eta_1 \partial^{(1)} \right)
 \left( \eta_2 \partial^{(2)} \right) \cdots & &  \\
 \cdots \left( \eta_{i-1} \partial^{(i-1)} \right)
 \left( \eta_{i+1} \partial^{(i)} \right) \cdots
 \left( \eta_{p+1} \partial^{(p)} \right) & & \nonumber\\
 F_p (\phi, \phi, \cdots, \phi) & & \nonumber
 \end{eqnarray}

\noindent
Now we take advantage of the identity (\ref{A.1}). It allows
us to make the replacements
\[
\partial^{(1)} \rightarrow - \partial^{(0)},
\partial^{(2)} \rightarrow - \partial^{(1)},
\cdots, \partial^{(i-1)} \rightarrow
- \partial^{(i-2)}.
\]

\noindent
Thus
\begin{eqnarray}
\label{B.6}
 D_2 \left( \eta_1, \cdots, \eta_p \right) =
 - \sum\limits^{p+1}_{i=1}
 \left( \eta_1 \partial^{(0)} \right)
 \left( \eta_2 \partial^{(1)} \right) \cdots
 \left( \eta_{i-1} \partial^{(i-2)} \right)
 \left( \eta_{i+1} \partial^{(i)} \right) & & \nonumber \\
 \cdots \left( \eta_{p+1} \partial^{(p)} \right)
 F_p (\phi,  \cdots, \phi) \, \, \, \, \,  & &
 \end{eqnarray}

\noindent
and therefore $D_2 + D^{p-1}_1 = 0$, i.e. $D = D^p_1$ :
\begin{eqnarray}
\label{B.7}
\left( \delta F_p \right) \left( \phi, \phi + \eta_1,
\cdots, \phi + \eta_{p+1} \right) \hspace*{3cm} & & \\
= \eta^{a_1}_1 \eta^{a_2}_2 \cdots \eta^{a_{p+1}}_{p+1}
\, \, \, \partial^{(0)}_{a_1} \partial^{(1)}_{a_2} \cdots
   \partial^{(p)}_{a_{p+1}} F_p (\phi, \cdots, \phi)
& & \nonumber
\end{eqnarray}

\noindent
The antisymmetrized components of this tensor are
\begin{eqnarray}
\label{B.8}
& &  \partial^{(0)}_{\left[ a_1 \right.} \partial^{(1)}_{a_2}
\cdots   \partial^{(p)}_{a_{\left. p+1 \right]}}
F_p  (\phi, \cdots, \phi)   \nonumber \\
& = & ( \partial^{(0)}
            + \partial^{(1)} + \cdots + \partial^{(p)} )
_{\left[ a_1 \right.}
  \partial^{(1)}_{a_2} \cdots  \partial^{(p)}_{a_{\left. p+1 \right]}}
F_p (\phi, \cdots, \phi)   \nonumber \\
& = & \partial_{\, \, \left[ a_1 \right.}
   {\rm Cl} \left( F_p \right)
 _{\left. a_2 a_3 \cdots a_{p+1} \right]}   \\
& = & \left( d \,{\rm Cl} ( F_p) \right)_{a_1 \cdots
a_{p+1}}  \nonumber
\end{eqnarray}

\noindent
The derivative
\begin{equation}
\label{B.9}
\partial_a \equiv \partial^{(0)}_a + \partial^{(1)}_a +
\cdots + \partial^{(p)}
\end{equation}

\noindent
acts on all arguments of $F_p$, but only $\partial^{(0)}_a$ survives
the antisymmetrization. Eq.~(\ref{B.7}) with (\ref{B.8}) proves
our claim (\ref{B.1}).\\

\noindent
Next we derive eq.~(\ref{2.24})
\begin{equation}
\label{B.10}
{\rm Cl} \left( {\cal L}_p F_p \right) = l_h {\rm Cl}
\left( F_p \right) ,
\end{equation}

\noindent
where
\begin{equation}
\label{B.11}
{\cal L}_p = \sum\limits^p_{i=0} h^a \left( \phi_i \right)
\partial^{(i)}_a  .
\end{equation}

\noindent
This time we have to expand
\begin{equation}
\label{B.12}
{\rm E} \left( \eta_1, \cdots, \eta_p \right) \equiv
\left( {\cal L}_p F_p \right) \left( \phi, \phi + \eta_1,
\cdots, \phi + \eta_p \right) .
\end{equation}

\noindent
Formally setting $\eta_0 \equiv 0$, ${\cal L}_p$, with the
arguments shifted, reads
\begin{eqnarray}
\label{B.13}
\sum\limits^p_{i=0} h^a \left( \phi + \eta_i \right)
\partial^{(i)}_a \hspace*{2.8cm} \\
= h^a ( \phi ) \sum\limits^p_{i=0} \partial^{(i)}_a
+ \sum\limits^p_{i=1} \eta^b_i \, \partial_b \, h^a
( \phi ) \, \,\, \partial^{(i)}_a + \cdots\nonumber
\end{eqnarray}

\noindent
Inserting (\ref{B.13}) into (\ref{B.12}) and expanding $F_p$
      itself leads
to ${\rm E} = {\rm E}_1 + {\rm E}_2$ with
\begin{eqnarray}
{\rm E}_1 \left( \eta_1, \cdots, \eta_p \right) =
h^a (\phi) \sum\limits^p_{i=0} \partial^{(i)}_a \, \, \exp
\left[ \sum\limits^p_{j=1} \left( \eta_j \partial^{(j)} \right) \right]
F_p \, ( \phi, \cdots, \phi) \label{B.14}  \\
& & \nonumber \\
{\rm E}_2 \left( \eta_1, \cdots, \eta_p \right) =
\partial_b h^a (\phi) \sum\limits^p_{i=1} \eta^b_i
\partial^{(i)}_a \, \, \exp
\left[ \sum\limits^p_{j=1} \left( \eta_j \partial^{(j)} \right) \right]
  F_p ( \phi, \cdots, \phi)   \label{B.15}
\end{eqnarray}

\noindent
Let us first look at ${\rm E}_1$, (\ref{B.14}). We would like to
conclude that the term with $p$ different $\eta$'s is the
only one which survives the expansion, and that there are no terms
with a smaller number of $\eta$'s. Because of the presence
of the operator $\sum_i \partial^{(i)}_a$, we cannot
apply eq.~(\ref{A.3}) directly, but it is easy to derive
an appropriate generalization of it. To every
$F_p \in \bigwedge^p_{\rm MS}$ we can associate a new
$\widetilde{F}_p \in \bigwedge^p_{\rm MS}$ by
shifting all $p+1$ arguments by a constant vector $v$:
\begin{eqnarray}
\label{B.16}
\widetilde{F}_p \left( \phi_0, \cdots, \phi_p \right) & = &
F_p \left( \phi_0 + v, \cdots, \phi_p + v \right) \\
& = & \exp \left[ v^b \sum\limits^p_{i=0}
 \partial^{(i)}_b \right] F_p
\left( \phi_0, \phi_1, \cdots, \phi_p \right) \nonumber
\end{eqnarray}

\noindent
Now we apply (\ref{A.3}) to $\widetilde{F}_p$ and take the
derivative $\frac{\partial}{\partial v^b}$ at $v = 0$.
This yields the desired relation:
\begin{equation}
\label{B.17}
\left( \sum\limits^p_{i=0} \partial^{(i)}_b \right) \, \,
\partial^{(l_1)}_{a_1} \cdots \partial^{(l_r)}_{a_r} F_p
( \phi, \phi, \cdots, \phi) = 0
\end{equation}

\noindent
It holds for all $r = 1, 2, \cdots, p-1$ provided all the
$l_i$'s are different from each other. Therefore, applying
(\ref{B.17}) to (\ref{B.14}) leads to
\begin{equation}
\label{B.18}
{\rm E}_1 \left( \eta_1, \cdots, \eta_p \right) =
\eta^{a_1}_1 \cdots \eta^{a_p}_p \, \, h^a
( \phi ) \partial_a \, \,
\partial^{(1)}_{a_1} \cdots \partial^{(p)}_{a_p}
F_p ( \phi, \cdots, \phi)
\end{equation}

\noindent
where also eq.~(\ref{B.9}) has been used.

Let us now turn to ${\rm E}_2$, (\ref{B.15}), which may be
represented as
\begin{eqnarray}
\label{B.19}
{\rm E}_2 \left( \eta_1, \cdots, \eta_p \right) =
\partial_b h^a (\phi) \sum\limits^p_{i=1}
\eta^b_i \frac{\partial}{\partial \eta^a_i} \\
\exp \left[ \sum\limits^p_{j=1} \eta^c_j
 \partial^{(j)}_c \right] F_p ( \phi, \cdots, \phi)
\nonumber
\end{eqnarray}

\noindent
Because the operator $\sum \eta \frac{\partial}{\partial \eta}$
can be applied after the $\partial^{(j)}$--derivatives have
been taken and the coincidence limit has been performed,
eq.~(\ref{A.3}) immediately implies that only the term with
$p$ factors of $\eta$ survives from $\exp [ \cdots ] \, F_p$
$( \phi, \cdots, \phi)$. Therefore
\begin{eqnarray}
\label{B.20}
{\rm E}_2 \left( \eta_1, \cdots, \eta_p \right) =
\eta^{a_1}_1 \cdots \eta^{a_p}_p
\sum\limits^p_{i=1} \partial_{a_i} h^b (\phi)\nonumber \\
\partial^{(1)}_{a_1} \cdots \partial^{(i-1)}_{a_{i-1}}
\, \partial^{(i)}_{b} \, \partial^{(i+1)}_{a_{i+1}} \cdots
\partial^{(p)}_{a_p} \, F_p \, (\phi, \cdots, \phi)
\end{eqnarray}

\noindent
This argument justifies also the truncation of the series
(\ref{B.13}) at order $\eta$. Combining (\ref{B.18}) with
(\ref{B.20}) leads to
\begin{eqnarray}
\label{B.21}
\left( {\cal L}_p F_p \right) \left( \phi, \phi + \eta_1,
\cdots, \phi + \eta_p \right)  \nonumber \\
= \eta^{a_1}_1 \cdots \eta^{a_p}_p
\left[ h^b (\phi) \partial_b \alpha_{a_1  \cdots a_p} \right.
( \phi ) \nonumber \\
+ \sum\limits^p_{i=1} \partial_{a_i} h^b
\left. ( \phi ) \alpha_{a_1 \cdots b \cdots a_p} (\phi) \right]
\end{eqnarray}

\noindent
with
\begin{eqnarray}
\label{B.22}
\alpha_{a_1 \cdots a_p} (\phi) & \equiv &
\partial^{(1)}_{a_1} \cdots \partial^{(p)}_{a_p} \, F_p\,
(\phi, \cdots, \phi ) \\
& = & \left[ {\rm Cl} \left( F_p \right) \right]
_{a_1 \cdots a_p}  (\phi) \nonumber
\end{eqnarray}

\noindent
Recalling (\ref{2.25}) we see that (\ref{B.21}) is exactly
what we wanted to prove, viz. eq.~(\ref{B.10}).

\end{document}